\documentclass[journal,twoside]{ieeeconf}
\usepackage[utf8]{inputenc}
\usepackage{tikz}
\usepackage{circuitikz}
\usepackage{amsmath,amssymb,graphicx}%,bbm,color,amstext,wasysym,flushend,parskip}

\usepackage{caption}
\definecolor{grey}{rgb}{0.6,0.6,0.6}
\definecolor{lightgray}{rgb}{0.97,.99,0.99}
 \newcommand{\dbar}{d\hspace*{-0.1em}\bar{}\hspace*{0.2em}}
\definecolor{llgrey}{rgb}{0.9,0.9,0.9}
\definecolor{lgrey}{rgb}{0.6,0.6,0.6}
\definecolor{lred}{rgb}{0.9,0.7,0.7}

\definecolor{dblue}{rgb}{0,0.5,.7}

\newtheorem{prop}{Proposition}

\newtheorem{remark}{Remark}

\newcommand{\trace}{{\rm Tr\,}}

\def\spacingset#1{\def\baselinestretch{#1}\small\normalsize}
\spacingset{1}

\newcommand{\Expect}{\mathbb{E}}
\newcommand{\ud}{d}

\newcommand{\mR}{\mathbb{R}}

\begin{document}
\title
%{Power transfer in Stochastic Thermodynamics}
{A matching principle for power transfer\\ in Stochastic Thermodynamics}
%Impedance matching in Stochastic Thermodynamics}

\author{Olga Movilla Miangolarra$^\star$, Amirhossein Taghvaei$^\dagger$, and Tryphon T. Georgiou$^\star$
\thanks{$^\star$Mechanical and Aerospace Engineering, University of California, Irvine, CA 92697, USA;
omovilla@uci.edu, tryphon@uci.edu}
\thanks{$^\dagger$Aeronautics and Astronautics Department, University of Washington, Seattle, Washington 98195, USA; amirtag@uw.edu}}

\maketitle

\begin{abstract}
Gradients in temperature and particle concentration fuel many processes in the physical and biological world. In the present work we study a thermodynamic engine powered by anisotropic thermal excitation (that may be due to e.g., a temperature gradient), and draw parallels with the well-known principle of impedance matching in circuit theory, where for maximal power transfer, the load voltage needs to be half of that of the supplying power source. We maximize power output of the thermodynamic engine at steady-state and show that the optimal reactive force is precisely half of that supplied by the anisotropy.
\end{abstract}


\section{Introduction}
Imagine a windmill, with blades at an angle with respect to wind velocity, and the wind coming straight at it.
The windmill draws power for rotational speeds $\omega\in(0,
\omega_{\rm max})$; clearly, if it is not rotating no power is drawn, and if it rotates too fast in the direction of the wind, power is delivered instead of drawn.
At what angular velocity is the power drawn maximal? This depends on the geometry of the blades and is not the subject of our paper. However, it is intuitively clear that the ``sweet spot'' is somewhere in the middle, where the product of torque applied by the wind times the angular velocity, hence power, is maximal. This helps highlight a general principle.

In some detail, and in order to draw parallels later on, assume a first-order approximation for the supplied torque by the wind $\tau_S-\omega R$. The dynamics of the angular velocity of the windmill obey 
$J\dot \omega = \tau_S-\omega R-\tau_L$,
where $\tau_L$ represents the torque of the load. At steady-state, 
\begin{equation}\label{eq:matching}
    \omega=(\tau_S-\tau_L)/R
\end{equation}
and the power drawn is $P=\omega\tau_L$. Clearly, the power is positive for $\omega\in(0,\omega_{\max}=\tau_S/R)$, and is maximal  for $\omega^*=\omega_{\max}/2$ with an optimal load torque
\[
\tau^*_L=\tau_S/2.
\]
Thus, as expected, the ``sweet spot''  is in the middle.

\begin{figure}
    \centering
\begin{circuitikz} \draw
(0,2) to[battery] (0,0) to (4,0) to[R=$R_L$] (4,2) to [R=$R_S$] (0,2);
\node (A) at  (0.3,1.7) {$V_S$};
\node (B) at  (4.3,1.7) {$V_L$};
\node (C) at  (-0.3,1.7) {$+$};
\node (D) at  (3.7,1.7) {$+$};
\draw (2,0) to (2,-.3);
\draw (1.8,-.3) to (2.2,-.3);
\draw (1.9,-.4) to (2.1,-.4);
\node (G) at  (3.2,2.3) {$i$};
\node (E) at  (3,2) {};
\node (F) at  (3.5,2) {};
\draw [->] (E) edge (F);
\end{circuitikz}
    \caption{Impedance matching example.}
\label{fig:electricalcircuit}
\end{figure}
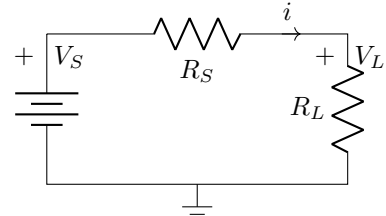

This same principle is often referred to as impedance matching in circuit theory. We briefly point to a textbook example. Consider a voltage source $V_S$ with internal resistance $R_S$ (that includes that of the transmission line) and a load with resistance $R_L$ as in Figure~\ref{fig:electricalcircuit}.
Then, the current is 
\[
i=(V_S-V_L)/R_S,
\]
where $V_L=i R_L$, and the power drawn is $P=i V_L$.
The power is maximal when the load resistance matches that of the source, $R_L=R_S$, equivalently, it is half of the total $R_L=(R_S+R_L)/2$.
Viewing the load as reacting by producing voltage $V_L$, the power is maximal (irrespective of $R_S$) when 
\[
V_L^*=V_S/2.
\]

The purpose of the present work is to point to a similar principle in
nonequilibrium thermodynamics. We consider 
a thermodynamic ensemble of particles subject to anisotropic fluctuations along different degrees of freedom, and we study the problem of maximizing power output at a {\em nonequilibrium steady-state} (NESS). The NESS can be pictured as a whirlwind, with the thermal anisotropy constituting the power source that sustains the circulatory steady-state current. The optimal load, effected by 
externally actuated non-conservative forces,  turns out to be half of that supplied by the anisotropy (cf.\ \eqref{eq:mainresult}).

In Section \ref{sec:BrownianGyrator} we present a canonical example of a system subject to anisotropic temperatures. Then, Sections \ref{sec:Gaussian} and \ref{sec:general}  develop the matching principle in the case where the potential is quadratic, and in general, respectively.

\section{The Brownian Gyrator}\label{sec:BrownianGyrator}
We consider overdamped Brownian particles with two degrees of freedom in an anisotropic heat bath, and subject to a quadratic potential. The location $X_t\in \mathbb R^2$ of the particles obeys
the Langevin dynamics
\begin{align}\label{eq:2dlangevin}
	d X_t &= \frac{1}{\gamma}\left({f(X_t)}-\nabla {U(X_t)}\right) d t + \sqrt{\frac{2k_BT}{\gamma}} d B_t,
\end{align}
with the force term consisting of a non-conservative component 
{$f(X_t)$} and the gradient of a potential 
{\begin{equation*}
U(X_t)=\frac{1}{2} X_t^\prime\, K_c \,X_t,
\end{equation*}}with {$K_c$} a positive definite $2\times 2$ matrix and $'$ denoting transpose. Throughout, $k_B$ is the Boltzmann constant,
$\gamma$ the friction coefficient, $\{B_t\}_{t\geq 0}$ denotes a $2$-dimensional standard Brownian motion (with mean zero and covariance the identity) and
$T={\rm diag}(T_1,\;T_2)$ a diagonal matrix with entries 
the temperature of thermal excitation along each of the two degrees of freedom.

The probability density function  of the state $X_t$ in \eqref{eq:2dlangevin}, denoted by\footnote{We use the notation $x$ for a vector in $\mathbb R^2$, and $X_t$ for the vector-valued stochastic process of position.} $\rho(t,x)$, with $x\in\mathbb R^2$,
% $p(x)\in P_2(\mR^n)$ 
constitutes the {\em state of the thermodynamic system} and represents the ensemble of particles; it satisfies the 
Fokker-Planck 
equation \cite{Handel07stochasticcalculus}
\begin{align}\label{eq:fokker}
\partial_t\rho  + \nabla \cdot\left(\rho v\right)=0,
\end{align}
with
$$v=-\frac{1}{\gamma}(\nabla U-f  + k_B T \nabla \log (\rho)),$$The non-conservative term $f$, when present, is assumed to be divergence-free with respect to $\rho$, in that $\nabla \cdot (f\rho)=0$. Thus, it represents a force that is not due to a potential energy and has no effect on the evolution of the state of the system.

When the initial state is Gaussian with mean zero and covariance $\Sigma_0$, denoted $\mathcal N(0,\Sigma_0)$, and $f$ is a linear function of $X_t$, then $\rho$ remains Gaussian over time with  mean zero and
covariance that satisfies the Lyapunov equation 
\begin{equation}\label{eq:Lyapunov}
    \gamma \dot\Sigma(t)=-{K_c}\Sigma(t)-\Sigma(t) {K_c}+2k_BT,
\end{equation}
that in this case consitutes the system dynamics.
The non-conservative force  is of the form
$f=\Omega\Sigma^{-1}(t)X_t$, with $\Omega$ a skew-symmetric matrix; it plays no role in \eqref{eq:Lyapunov} (i.e., it cancels out) due to the skew symmetry $\Omega+\Omega^\prime=0$.

At any point in time, the total energy of the system is $E=\int U(x) \rho(t,x) dx$. The system exchanges energy with the environment either through the external non-conservative forcing (work differential in \eqref{eq:work-p}), or through heat transfer to and from the two thermal baths (heat differential in \eqref{eq:heat-p}). Specifically, the first of these contributions reads~\cite{sekimoto2010stochastic, seifert2012stochastic},
\begin{equation}\label{eq:work-p}
    \dbar W = \Expect[f' \circ dX_t],
\end{equation}
and represents work\footnote{Here, ``$\circ$" denotes Stratonovich integration, while $\dbar$ indicates an imperfect differential, where its integral depends on the chosen path and not only on the end points.}.
Upon writing this expression in It$\rm\hat{o}$ form and using \eqref{eq:2dlangevin}, we obtain
\begin{align*}
    \dbar W =&\frac{1}{\gamma}\int \Big(f'(-\nabla U+f)+k_B\trace[{\rm Jac}(f)T]\Big)\rho dxdt\\=& 
\int f'v\rho dxdt,
\end{align*}
where we have used the It$\rm\hat{o}$ rule and integrated by parts. Here,  ${\rm Jac}(f)$ denotes the Jacobian of $f$.

On the other hand, the heat uptake from the thermal baths is~\cite{sekimoto2010stochastic, seifert2012stochastic}
\begin{equation}\label{eq:heat-p}
    \dbar Q =\dbar Q_1+\dbar Q_2=  \iint  (\nabla U-f)' v\rho\,dxdt,
\end{equation}
which can be split into the contributions coming from the different heat baths, $\dbar Q_1$ and $\dbar Q_2$.
Combining \eqref{eq:work-p} and \eqref{eq:heat-p}, we have the first law of thermodynamics,
\begin{equation}\label{eq:1st}
    dE=\dbar W+\dbar Q,
\end{equation}
where the differential of internal energy is the sum of the two contributions.

\subsection*{Steady-state analysis}

Let us assume momentarily that $f=0$. Since the potential is quadratic with $K_c$ constant and positive definite, the system eventually reaches a steady-state distribution; this is Gaussian $\mathcal N(0,\Sigma_{ss})$ with (steady-state) covariance satisfying the algebraic Lyapunov equation
\begin{equation}\label{eq:ARE}
K_c\Sigma_{ss}+\Sigma_{ss} K_c=2k_BT.  
\end{equation}
The solution to \eqref{eq:ARE} is unique and can be expressed in integral form as
\begin{equation}\label{eq:ARE-sol}
\Sigma_{ss} = 2k_B\int_0^\infty e^{-\tau K_c} Te^{-\tau K_c}d\tau
 =2k_B \mathcal L_{K_c}(T),   
\end{equation}
where, for future reference, we define the linear operator
\[
X \mapsto \mathcal L_{A}(X) :=  \int_0^\infty   e^{-\tau {A}} Xe^{-\tau {A}}d\tau
\]
that depends on the positive definite matrix ${A}$.

The velocity field becomes
\begin{align}\label{eq:vel-gaussian}
  \nonumber  v(x)&=\frac{1}{\gamma}(-K_cx+k_BT\Sigma^{-1}_{ss}x)\\&=-\frac{1}{2\gamma}(K_c\Sigma_{ss}-\Sigma_{ss}K_c)\Sigma_{ss}^{-1}x,
\end{align}
where we have used the algebraic Lyapunov equation \eqref{eq:ARE}.
Even if the system is at steady-state, the probability current $v\rho$ does not need to vanish, in general. A non-vanishing probability current induces a heat flow between the thermal baths, with $\dbar Q_1=-\dbar Q_2$~\cite{BGyrator2007first}.
When it vanishes ($v\rho=0$), a condition known as {\em detailed balance} in the physics literature, the steady-state is an {\em equilibrium state}~\cite{seifert2012stochastic}.

It is seen that in order to ensure detailed balance, $K_c$ and $\Sigma_{ss}$ must commute.
In view of \eqref{eq:ARE}, also \eqref{eq:ARE-sol}, this only happens when $T$ and $K_c$ commute; i.e., this happens when $K_c$ is diagonal, since $T$ is already diagonal. In that case, $\Sigma_{ss} = k_BTK_c^{-1}$ is also diagonal and results in zero probability current according to~\eqref{eq:vel-gaussian}. Moreover, in that case, heat cannot transfer between the degrees of freedom.
On the other hand, when $K_c$ and $T$ do not commute, detailed balance breaks down and a non-vanishing (stationary) probability current materializes leading to a NESS with non-vanishing heat transfer between the heat baths \cite{BGyrator2017experimental}. 

The system described here is known as the Brownian Gyrator. Since its conception by Filliger and Reimann in 2007 \cite{BGyrator2007first}, it has been thoroughly studied, mostly at steady-state. Several works focused on the curl-carrying probability current that mediates a heat transfer between heat baths and the torque that it generates \cite{BGyrator2007first,BGyrator2013CilibertoExperim,BGyrator2013dotsenko,BGyrator2017electrical,BGyrator2017experimental}. Other works have studied optimal transitioning between states  \cite{baldassarri2020engineered}, maximum work extraction through periodic variation of the potential function \cite{EnergyHarvesting2021,miangolarra2022geometry}, the role of information flow  \cite{BGinfo,Loos_2020nonreciprocal}, the effect of external forces \cite{constantforceBG},  strong coupling limits \cite{Imparato2017BG}, the control relevance of anharmonic potentials \cite{gyratingcharacnonharmonic}, gyration for underdamped mesoscopic systems \cite{inertiaBG}, the relevance of non-Markovian noise \cite{NonmarkovianBG} and the use of active reservoirs \cite{activereservoirBG}.

\begin{figure}
    \centering
\includegraphics[width=0.475\textwidth]{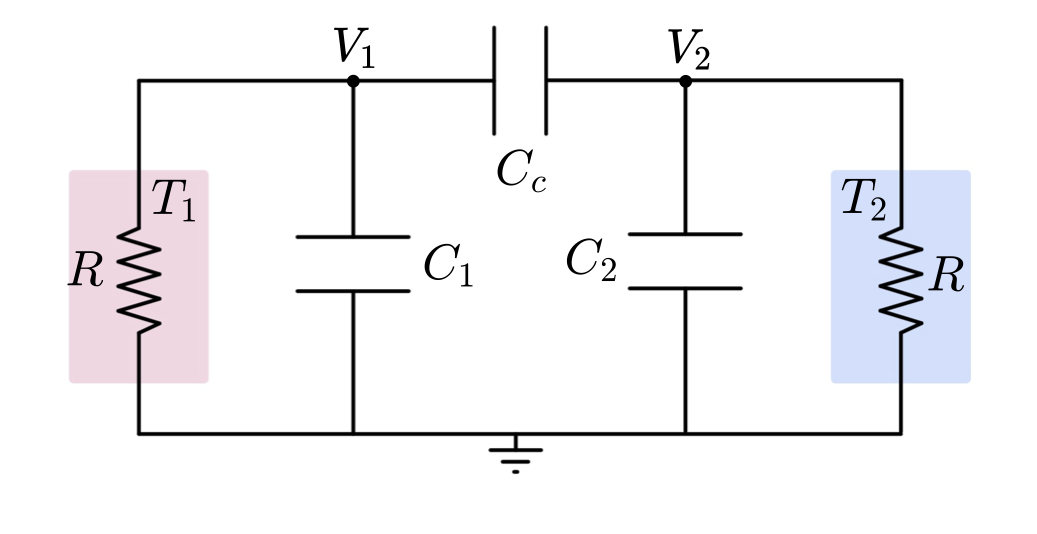}
    \caption{Electrical emboddiment of the Brownian gyrator.}
    \label{fig:circuit}
\end{figure}

Experimental realizations of \eqref{eq:2dlangevin} have been based on several different physical embodiments \cite{BGyrator2017experimental,BGyrator2017electrical,BGyrator2013CilibertoExperim,Bgyrator2013ciliberto, abdoli2022tunable}. In particular, \cite{BGyrator2017experimental,abdoli2022tunable} are based on colloidal particles suspended in a viscous medium while the
anisotropy in stochastic excitation is induced through an electromagnetic field.
Another embodiment is based on the electric circuit of 
Figure \ref{fig:circuit}.
Here, the two degrees of freedom are the charges in the capacitors $C_1$ and $C_2$, and the two resistors, in contact with heat baths of different temperatures, generate Johnson-Nyquist fluctuating currents \cite{Bgyrator2013ciliberto,BGyrator2013CilibertoExperim,BGyrator2017electrical}.

Specifically, the equations of motion of the system in Figure \ref{fig:circuit} can be written as \cite{BGyrator2017electrical}
$$
C dV_t=-\frac{1}{R}V_tdt+\sqrt{\frac{2 k_B T}{R}}dB_t,
$$
where $V_t=[V_1(t),\ V_2(t)]'$ are voltages across the two capacitors, $C$ is the matrix of capacitances
\begin{align*}
 &C=\left[\begin{array}{cc}
    C_1+C_c &-C_c  \\
    -C_c & C_2+C_c
\end{array}\right],
\end{align*}
and $\sqrt{2k_BT/R}\,dB_t$ models the
 Johnson-Nyquist noise \cite{nyquist} at the two resistors for temperatures $T_1$ and $T_2$ ($T={\rm diag}(T_1,T_2)$, as before).
The state of this electrical-thermal system can alternatively be described in terms of the charges $q_1(t)$ and $q_2(t)$ at capacitances $C_1$ and $C_2$, since the vector of charges is $q_t=C V_t$. 
Then, the charges satisfy
\begin{equation}\label{eq:langevin-electric}
\ud q_t=-\frac{1}{R}C^{-1} q_t\ud t+\sqrt{\frac{2 k_B T}{R}}dB_t.
\end{equation}

Let $U(q)=\frac{1}{2}q'C^{-1}q$  be the 
(potential) energy stored in the system of capacitances ($C_1,C_2,C_c$). The first term in the right-hand-side of \eqref{eq:langevin-electric} is precisely the negative gradient $-\nabla U(q_t)/R\ud t$. Hence, equation \eqref{eq:langevin-electric} 
represents a two-dimensional overdamped Langevin system \eqref{eq:2dlangevin} with non-conservative forces absent, and $R$ playing the role of the friction coefficient.
It follows that the distribution of charges $\rho(t,q_t)$ satisfies a Fokker-Planck equation \eqref{eq:fokker}, with velocity field $v$ being replaced by a corresponding current field (see equation \eqref{eq:boldi}).

\section{Maximum power extraction from the Brownian Gyrator}\label{sec:Gaussian}

In the present section, we take the next natural step and consider work extraction from the resulting circulating current. 
We explore
the coupling of the natural gyrating motion with an external actuation, for the purpose of extracting
work from the anisotropy of the temperature field. We restrict ourselves to non-conservative actuation (divergence-free), in order to maintain the system at steady-state\footnote{The analysis herein differs from instances where time-varying actuation is used to extract work when a system is in contact with a single heat bath with time-varying temperature profile \cite{overdampedseifert,fu2020maximal,olga2021lowfric}.}. Earlier studies on maximizing work output for this type of anisotropic, stationary systems~\cite{pietzonka2018BGeffic,abdoli2022tunable,NCbrownianG2022} were particular to linear two-dimensional systems.
Our contribution lies in a general approach that applies to higher dimensions and nonlinear forces (see Section \ref{sec:general}), while bringing to light the aforementioned matching principle, that the velocity ensuring maximal power transfer is precisely at the midpoint of the power producing range of velocity values.
% \black

Let us picture the steady-state of the Brownian gyrator (the system described by \eqref{eq:2dlangevin} subject to a fixed quadratic potential with $f=0$) as a vortex of swirling particles around the origin, which originates from the non-zero probability current at steady-state. Now imagine yourself sailing around the origin, propelled by the wind of particles, going in circles at some velocity slower than the particles, yet non-zero, slowed-down by external forces applied to the boat for the purpose of extracting work. The work extracted would correspond to the force applied on the sails times the displaced distance, in a way that is analogous to the windmill example discussed in the introduction.

Indeed, one can use non-conservative actuation to implement such an interaction. Let us consider 
$
f=\Omega\Sigma_{ss}^{-1}\,X_t,%=f_c\, \Omega\Sigma_{ss}^{-1}X_t, \ \mbox{ where } \ f_c\in \mathbb R \ \mbox{ and } \ \Omega=\left[\begin{array}{cc}
 %  0  & 1 \\
  % -1  &  0
%\end{array}\right].
$
 which represents the force on the sail that, due to its skew-symmetry, does not alter the stationary distribution of the state of the system (characterized by $\Sigma_{ss}$), but does affect the mean velocity $v$.
Thus, we write
 $$
    v=(\mathbf{f}_S-\textbf{f}_L)/\gamma,
$$
where we have defined the source and load forces  by\footnote{The usage of boldface font aims to highlight analogies in the role played by different physical quantities, e.g., between $\mathbf f_S$ and $\mathbf V_S$.} 
\begin{subequations}\label{eq:forces}
\begin{align}\label{eq:fs}
    \mathbf{f}_S&:=-K_cX_t+ k_BT \Sigma_{ss}^{-1}X_t\\\label{eq:fl}
    \mathbf{f}_L&:=-\Omega\Sigma_{ss}^{-1}X_t.
\end{align}
\end{subequations}
{In this, as in \eqref{eq:vel-gaussian}, we can also write
$$\mathbf{f}_S=-\Omega_c \Sigma_{ss}^{-1}X_t,
$$
for a skew symmetric matrix $$\Omega_c:=\frac{1}{2}(K_c\Sigma_{ss}-\Sigma_{ss}K_c).$$}The work output is now expressed as
{\begin{align*}\label{eq:workrateNC}
    -\dbar W &= \Expect\big[\mathbf{f}_L'(\mathbf f_S-\textbf{f}_L)/\gamma\big]dt\\\nonumber &=\frac{1}{\gamma}\Expect\big[-X_t'\Sigma_{ss}^{-1}\Omega'(\Omega-\Omega_c)\Sigma_{ss}^{-1}X_t\big]dt,
\end{align*}
and upon taking expectation we obtain that
\[
P=-\dbar W/dt=-\frac{1}{\gamma}
\trace[
(\Omega- \Omega_c) \Sigma_{ss}^{-1}\Omega^\prime
].
\]
}

Thus, the problem to maximize power $P$ is equivalent to the static problem
\begin{align*}
   \min \frac{1}{\gamma}
\trace[
(\Omega- \Omega_c) \Sigma_{ss}^{-1}\Omega^\prime
],
\end{align*}
where the minimum is taken over skew-symmetric $\Omega$'s. The first-order necessary condition for optimality is found by computing the first order variation of the cost and setting it to zero, this is,
\begin{equation}\label{eq:NCFONC}
  \trace[\Delta_\Omega M]=0,\ \mbox{ with } \ M=(2\Omega-\Omega_c)\Sigma^{-1}_{ss},
\end{equation}
and $\Delta_\Omega$ any skew-symmetric matrix. Therefore, the first-order necessary condition for optimality \eqref{eq:NCFONC} implies that $M$ must be symmetric. It follows that  optimal choice\footnote{Throughout, we superscribe $*$ to denote an optimal solution.} $\Omega^*$ is
\begin{equation*}
    \Omega^*=\frac12\Omega_c,
\end{equation*}
yielding $M=0$, for otherwise $M$ is not symmetric. Hence, we have established the following proposition.

\begin{prop}
The maximum steady-state power output through non-conservative (divergence-free) forcing for the linear system in \eqref{eq:2dlangevin},
\begin{equation}
    \max_{\mathbf{f}_L} \,\Expect[\mathbf{f}_L'(\mathbf{f}_S-\mathbf{f}_L)/\gamma],
\end{equation}
is obtained for 
\begin{equation}\label{eq:mainresult}
    \mathbf f_L^{\,*}=\mathbf f_S/2.
\end{equation}
\end{prop}
\begin{remark}
The maximum power output can be computed to be
\begin{align*}
 P^*&=\frac{1}{4\gamma}
\trace[
\Omega_c \Sigma_{ss}^{-1}\Omega_c^\prime
],
\end{align*}
which is quadratic in the source of power, i.e., the circulation in the velocity field $v$ \eqref{eq:vel-gaussian}.
Moreover, this expression for $P^*$ can be written explicitly in terms of the parameters of our system: first solve for the steady-state covariance $\Sigma_{ss}$ in \eqref{eq:ARE} (which is fairly simple for a two-dimensional problem), then use it write $\Omega_c$ and finally to write $P^*$. Explicit derivations in a particular two-dimensional case can be found in \cite{pietzonka2018BGeffic,NCbrownianG2022}.
\end{remark}

\begin{figure}
    \centering  \includegraphics[width=0.49\textwidth]{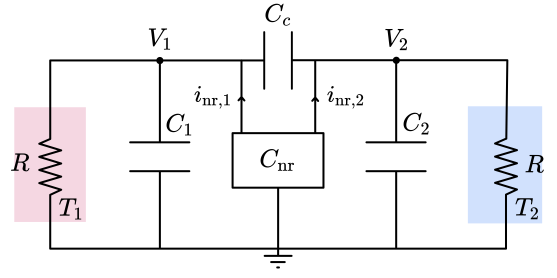}
    \caption{Circuit realization of a steady-state Brownian gyrator engine.}
    \label{fig:nc}
\end{figure}

Let us now return to a circuit theoretic embodiment. 
In order to generate a non-conservative force, a non-reciprocal capacitance is needed. To this end, we consider the circuit depicted in
%Envisioning an embodiment such as the one depicted in 
Figure \ref{fig:nc}, where we have introduced such a general (two-port) capacitance, with law
$$
i_{\rm nr}=\left[\begin{matrix}i_{{\rm nr},1} \\ i_{{\rm nr},2}\end{matrix}\right]=-C_{\rm nr} \frac{dV_t}{dt},
$$
and a capacitance matrix $C_{\rm nr}$ that is not necessarily symmetric (i.e., possibly non-reciprocal \cite{virtualpot2021Chang}).
The state of the system is the vector of charges $q_t$, as before, and satisfies \eqref{eq:langevin-electric} with $\hat C= C+C_{\rm nr}$ replacing  $C$. The ``force'' $-\hat C ^{-1}q_t$ can be split into conservative and non-conservative components,
$$
-\hat C^{-1}q_t=-(\hat C^{-1})^\prime C\hat C^{-1}q_t-(\hat C^{-1})^\prime C_{\rm nr}^\prime \hat C^{-1}q_t,
$$
with the non-conservative part $(\hat C^{-1})^\prime C_{\rm nr}^\prime \hat C^{-1}q_t=\Omega\Sigma_{ss}^{-1}q_t$.

As noted, the probability density of the system of charges follows the Fokker-Planck equation \eqref{eq:fokker}, with velocity field $v$ replaced by a current field (function of a vector\footnote{Following our earlier convention, $q_t$ represents the vector-valued stochastic process of charges, whereas $q$ represents a vector in $\mathbb R^2$.} $q$),
\begin{equation}\label{eq:boldi}
\mathbf i=(\mathbf{V}_S-\mathbf V_{L})/R,
\end{equation}
with
\begin{align*}
    \mathbf{V}_S&=-(\hat C^{-1})^\prime C\hat C^{-1}q+k_BT\Sigma^{-1}_{ss}q,\\
    \mathbf V_{L}&=-(\hat C^{-1})^\prime C_{\rm nr}^\prime \hat C^{-1}q,
\end{align*} 
and $q\in\mathbb R^2$.

The thermodynamic definition of power is consistent with the circuit theoretic viewpoint where $P=\Expect[V_t'\circ i_{\rm nr}],$ for $V_t=[V_1(t) \ V_2(t)]'=\hat C^{-1}q_t$. Specifically,
\begin{align*}
 Pdt&=\Expect[V_t'\circ i_{\rm nr} \, dt ]=-\Expect [V_t'\circ C_{\rm nr}dV_t]\\
 &=-\Expect [q_t'(\hat C^{-1})^\prime \circ C_{\rm nr} \hat C^{-1}dq_t]\\
 &= \Expect [\mathbf{V}_L' \circ dq_t],
\end{align*}
which, in analogy with \eqref{eq:work-p}, gives \[
-\dbar W=\Expect [\mathbf V_L'(\mathbf V_S-\mathbf V_L)/R]dt.
\]
Thus, maximum power is similarly obtained by a load voltage that halves the source voltage fueled by the anisotropy in temperature of the two Johnson-Nyquist resistors, i.e.  
$$
\mathbf V_L^*=\mathbf V_S/2,
$$ 
akin to the standard impedance matching problem.

\begin{remark} A circuit such as the one depicted in Figure \ref{fig:nc} can be realized through controlled feedback \cite{virtualpot2021Chang}. However, this requires external energy input which is unaccounted for, constituting the main disadvantage when attempting to extract work from a system through non-conservative forcing.
\end{remark}

\section{A general setting}\label{sec:general}

The previous results apply more generally to a linear system with $n$-degrees of freedom. We can do even better and consider an $n$-dimensional general case where equation  \eqref{eq:2dlangevin} holds  
with $X_t\in \mR^n$, $T$ a diagonal $n\times n$ matrix with the temperatures of the ambient heat baths as entries, 
 $B_t$ an $n$-dimensional standard Brownian motion, and each degree of freedom subjected to a general non-linear force $f_i-\partial_{x_i} U$.
The probability density function  $\rho(t,x)$ of the process $X_t$ satisfies the Fokker-Planck equation \eqref{eq:fokker}, with $v=(\mathbf f_S-\mathbf f_L)/\gamma$ and
 \begin{align*}
    \mathbf f_S&:=-\nabla U  - k_B T \nabla \log (\rho)\\\mathbf f_L&:=-f,
\end{align*}
where the latter is divergence-free with respect to $\rho$ as before, i.e., $\nabla \cdot (\mathbf f_L\rho)=0$.

Let $\rho_{ss}$ be an admissible steady-state, i.e. such that  
there exists a unique time-independent potential $U_c$
that renders $\rho_{ss}$ stationary. Specifically, the potential and $\rho_{ss}$ must satisfy
\begin{equation}\label{eq:ss}
    \nabla\cdot((\mathbf f_S-\mathbf f_L) \rho_{ss}/\gamma)=0,
\end{equation}
with
$$
\mathbf f_S=-\nabla U_c  - k_B T \nabla \log (\rho_{ss}).
$$
Thus, equivalently,
\begin{equation}\label{eq:poisson}
\nabla\cdot(\rho_{ss}\nabla U_c)=-\nabla\cdot(\rho_{ss}k_BT\nabla\log(\rho_{ss})).
\end{equation}
It is seen that $U_c$ is specified by the gradient part of $k_BT\nabla\log(\rho_{ss})$, in that $U_c$ must satisfy the Poisson equation \eqref{eq:poisson}. It is assumed that \eqref{eq:poisson} has a unique solution\footnote{ See \cite{miangolarra2023minimal} for sufficient conditions on $\rho_{ss}$ for uniqueness of solution.} for all admissible $\rho_{ss}$.

With the potential fixed at $U_c$ and the system at steady-state $\rho_{ss}$, the power output due to the non-conservative forcing is given by
\begin{align*}
    P=-\dbar W/dt =
\Expect[ \mathbf{f}_L'(\mathbf f_S-\mathbf f_L)/\gamma].
\end{align*}
The optimal  $\mathbf f_L$ that maximizes the power output must be such that the first order variation of $P$,
\begin{align*}
\int (\delta_{ \, \mathbf f_L})'(\mathbf f_S-2\mathbf f_L)\rho_{ss} dx/\gamma,
\end{align*}
vanishes for all admissible $\delta_{ \, \mathbf f_L}$. 
Given that $\delta_{ \, \mathbf f_L}$ must be a divergence-free field with respect to $\rho_{ss}$, we obtain that 
$\mathbf{f}_S-2\mathbf{f}_L$
must be of gradient form, from orthogonality.
Since both $\mathbf{f}_L$ and $\mathbf{f}_S$ must be divergence-free, the first by construction and the second  due to steady-state \eqref{eq:ss}, we obtain that 
the optimal $\mathbf f_L$ must reduce the divergence-free swirling motion of the particles (induced by $\mathbf{f}_S$) by half, as is stated in the following proposition.
\begin{prop}
The maximum steady-state power output through divergence-free, non-conservative forcing for the (non-linear) system \eqref{eq:2dlangevin} with $X_t\in\mathbb R^n$,
\begin{equation}
    \max_{\mathbf{f}_L} \,\Expect[\mathbf{f}_L'(\mathbf{f}_S-\mathbf{f}_L)/\gamma],
\end{equation}
is obtained by 
\begin{equation}
    \mathbf f_L^{\,*}=\mathbf f_S/2.
\end{equation}
\end{prop}

\begin{remark}
    The maximum power output is given by
\begin{align*}
P^*=&\frac{1}{4\gamma}\Expect[\mathbf{f}_S'\,\mathbf{f}_S],
\end{align*}
which is quadratic in the source force.
\end{remark}

\section{Conclusions}
The purpose of this work has been to present a matching principle for maximal power extraction in diverse systems, ranging from microscopic thermodynamic heat engines to windmills and electric circuits.
This principle holds in general under the assumption that the source has a linear response (e.g., voltage-current, force-velocity, etc.).
%and the existence of a steady-state (c.f. \eqref{eq:matching}).
This is precisely the case underlying the well-known impedance matching principle for power transfer in circuits, and the same principle is extrapolated % translated
to stochastic systems where the anisotropy in thermal fluctuations
constitutes the source of energy, as explained in the body of the paper. Interest in this principle stems from the significance of power harvesting mechanisms in the physical and biological world. Thus, it appears of great importance to investigate whether naturally occurring analogues, such as those driving bacterial flagella \cite{blair1995bacteria,wadhwa2022bacterial}, have evolved to display some form of optimality that reminisces the matching principle that we discussed herein. 

We would like to remark that impedance matching in classical network theory extends to dynamical loads \cite{anderson2013network,belevitch1968classical}.
Similarly, instances where the power source may not be Markov and where the power generated is consumed by a higher-order dynamical component, are relevant in the context of thermodynamics. Finally, a thermodynamic counterpart of maximal power transfer \cite{youla1967interpolation,helton1980distance}  is of interest, where suitably designed coupling may facilitate impedance matching between given thermodynamic components.

\appendices

\section*{Acknowledgments}
The research was supported in part by the AFOSR under grant FA9550-20-1-0029, and ARO under W911NF-22-1-0292. O.M.M was supported by ”la Caixa” Foundation (ID 100010434) with code LCF/BQ/AA20/11820047.

\bibliography{arXiv}
\bibliographystyle{IEEEtran}

\end{document}